\documentclass[reprint,superscriptaddress, amsmath,amssymb, aps,floatfix]{revtex4-2}
\usepackage{array}
\usepackage{verbatim}
\usepackage{graphicx}
\usepackage[utf8]{inputenc}
\usepackage{stmaryrd}
\usepackage[T1]{fontenc}
\usepackage{dcolumn}
\usepackage{bm}
\usepackage{upgreek}
\usepackage{siunitx} 
\usepackage[normalem]{ulem}
\usepackage{hyperref}
\begin{document}

\title{Realization of Very High Mobility InAs Quantum Wells on InP Substrates}

\author{Tyler Lindemann} 
 \affiliation{Department of Physics and Astronomy, Purdue University, West Lafayette, Indiana 47907, USA}
 \affiliation{Microsoft Quantum, West Lafayette, Indiana 47907, USA}

\author{Rojila Ghimire}
 \affiliation{Department of Physics and Astronomy, Purdue University, West Lafayette, Indiana 47907, USA}
 
\author{Alejandro Alcaraz Ramirez}
  \affiliation{Microsoft Quantum, West Lafayette, Indiana 47907, USA}

\author{Ahmad Azizimanesh}
  \affiliation{Microsoft Quantum, West Lafayette, Indiana 47907, USA}

\author{Sergei Gronin}
 \affiliation{Microsoft Quantum, West Lafayette, Indiana 47907, USA}

\author{Ray Kallaher}
 \affiliation{Microsoft Quantum, West Lafayette, Indiana 47907, USA}
 
\author{Michael J. Manfra}
\email{mmanfra@purdue.edu; b-mmanfra@microsoft.com}
 \affiliation{Department of Physics and Astronomy, Purdue University, West Lafayette, Indiana 47907, USA}
 \affiliation{Microsoft Quantum, West Lafayette, Indiana 47907, USA}
 \affiliation{School of Materials Engineering, Purdue University, West Lafayette, Indiana 47907, USA}
 \affiliation{Elmore Family School of Electrical and Computer Engineering, Purdue University, West Lafayette, Indiana 47907, USA}
 \affiliation{Purdue Quantum Science and Engineering Institute, West Lafayette, IN, 47907, USA}
\date{\today}

\begin{abstract}
Due to strong spin-orbit coupling, large Land{\'e
}{\it g}-factor, and a highly transparent interface, the two-dimensional electron gas (2DEG) in InAs quantum wells provides an ideal platform for the exploration of topological superconductivity in hybrid semiconductor-superconductor heterostructures. Reduction of disorder in the semiconductor remains a primary challenge to the disambiguation of subtle interaction effects. Here we demonstrate very high mobility InAs 2DEGs grown on insulating InP substrates. Utilizing a convex compositional grading profile in an InAlAs metamorphic buffer we demonstrate InAs quantum wells with thickness greater than 10~nm without need of Ga-containing cladding layers. We examine structural and electrical transport properties of these heterostructures. Advances in the design of the metamorphic InAlAs buffer result in low-temperature electron mobility exceeding $1.7\times10^6~\mathrm{cm^2\,V^{-1}\,s^{-1}}$ at 2DEG density \mbox{$\leq3.2\times10^{11}~\text{cm}^{-2}$, the highest value yet reported for this material system.}
\end{abstract}

\maketitle

\begin{table*}[!t]
\caption{\label{Table_1}%
Overview of sample parameters explored in this study: quantum well thicknesses, $d$, peak mobility, $\mu_{max}$, 2DEG density at peak mobility, $n_{2DEG}$ at $\mu_{max}$, and 2DEG density at zero gate bias, $n_{2DEG}$ at $V_{g} = 0 \text{ V}$. The listed data was measured using gated Hall bars oriented along the $[1\bar{1}0]$ crystallographic direction.
}
\begin{ruledtabular}
\begin{tabular}{ccccc} 
 Sample & $d$ (nm) & $\mu_{max}$ ($\mathrm{{10^{6}\,cm^2\,V^{-1}\,s^{-1}}}$) & $n_{2DEG}$ at $\mu_{max}$ ($10^{11}$~cm$^{-2}$) & $n_{2DEG}$ at $V_{g} = 0 \text{ V }$ ($10^{11}$~cm$^{-2}$)\\
 \hline
 A & 10 & 1.56 & 3.28 & 3.50\\ 
 B & 12 & 1.53 & 3.01 & 3.30\\ 
 C & 14 & 1.72 & 3.24 & 3.48 \\ 
\end{tabular}
\end{ruledtabular}
\label{Table_1}
\end{table*}

\section{Introduction}

Strong spin-orbit coupling, large Land\'e {\it g}-factor, and highly transparent interfaces make InAs a useful platform for the study of topological superconductivity. InAs is readily proximitized by epitaxially-grown superconductors in shallow quantum well heterostructures. Due to its relatively small lattice mismatch to InAs, GaSb has been an obvious choice for hosting InAs-containing heterostructures \cite{Kroemer2004}, where relatively thick InAs quantum wells can be grown without relaxation \cite{Thomas2018}. Integration of InAs quantum wells on antimonide-based heterostructures presents challenges, however, particularly in optimizing the mixed Sb-As interfaces, with work ongoing since Tuttle {\it et al}. in the 1980s \cite{Tuttle1989}. Arsenide-only active regions integrated on semi-insulating InP via a graded buffer layer (GBL) offer an advantage in this regard, as they do not rely on mixed-anion barriers. In addition, the superior insulating properties of InP substrates compared to GaSb offers important advantages for high-frequency device operation.

Due to the substantial compressive strain in InAs quantum wells grown on InP substrates, quantum well thickness is often limited below 10~nm. In these cases, interface roughness scattering and alloy scattering from the barriers remain significant sources of disorder. One solution to minimize the impact of these disorder sources is to increase the thickness of the InAs quantum well in order to minimize the extent of 2DEG's wavefunction into the barrier and minimize overlap with the interfaces to the barriers. In previous studies, linear step grading profiles for the GBL have commonly been used, usually in conjunction with compositional overshoot and a compositional ‘step-back’ layer to eliminate residual strain accumulated
in the buffer \cite{Shabani2014APL,Hatke2017APL,Ayers2015}. To reduce lattice mismatch and grow thicker InAs quantum wells, one needs to
grade to a higher indium composition to achieve a larger lattice parameter close to that of InAs. The use of compositional overshoot, however, becomes untenable at high indium-concentrations, as it can lead to unwanted parallel
conduction channels near the primary InAs quantum well. As an alternative approach, here we employ a convex exponential grading profile for the GBL, facilitating the growth of thicker, coherently-strained InAs quantum wells than was previously possible for InAs on InP without the use of Ga-containing cladding barriers.

To place our results in context, we draw comparisons with previous reports of high mobility InAs quantum wells grown on InP substrates. Hatke \textit{et al.}~\cite{Hatke2017APL} demonstrated high mobility in lattice-mismatched InAs/InP heterostructures using \(\mathrm{In}_{0.75}\mathrm{Ga}_{0.25}\mathrm{As}\)
cladding layers and \(\mathrm{In}_{0.75}\mathrm{Al}_{0.25}\mathrm{As}\) barriers. Their optimized structures reached a peak mobility of approximately
\(1.1\times10^{6}~\mathrm{cm^2\,V^{-1}\,s^{-1}}\) at 2DEG density $6.2\times 10^{11}$~cm$^{-2}$~\cite{Hatke2017APL}. More recently, Dempsey 
\textit{et al.} utilized strain compensation by reducing the In concentration in the InGaAs cladding layers to enable a \(16~\mathrm{nm}\)-wide InAs quantum well and reported a peak mobility of \(1.16\times10^{6}~\mathrm{cm^2\,V^{-1}\,s^{-1}}\) at an electron density of \(4.2\times10^{11}~\mathrm{cm^{-2}}\) \cite{Dempsey2025PRM}.
In our current work, replacing the Ga-based cladding layers with
\(\mathrm{In}_{0.875}\mathrm{Al}_{0.125}\mathrm{As}\) barriers on top of a convex GBL enables the growth of thicker InAs quantum wells. Among the measured samples,
a \(14~\mathrm{nm}\) quantum well exhibits the highest mobility with $\mu = 1.72\times10^{6}~\mathrm{cm^2\,V^{-1}\,s^{-1}}$,
at an electron density of
$n = 3.24\times10^{11}~\mathrm{cm^{-2}}$.
The enhanced mobility in the wider quantum well is attributed to the enhanced confinement of 2DEG wavefunction in the quantum well leading to the reduction of the interface roughness and alloy-disorder scattering~\cite{Dempsey2025PRM,benali2022metamorphic} and to reduced background charged impurity scattering.

\begin{figure*}[!t]
  \centering
  \begin{minipage}{\textwidth}
    \centering
    \includegraphics[width=\textwidth]{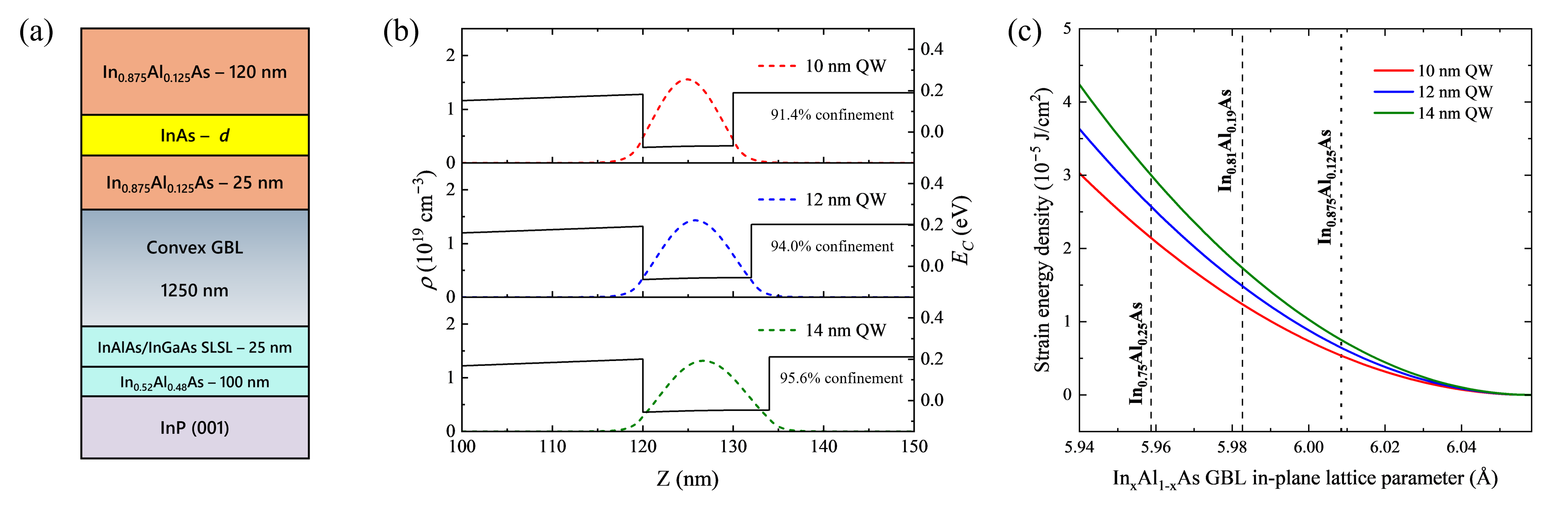}
    \caption{(a) The layer stack of the heterostructure used for the three samples examined in this study. The quantum well thickness, $d$, was varied as described in Table~\ref{Table_1}. (b) Self-consistent Schr\"{o}dinger-Poisson calculations of the conduction band edge and electron charge distribution of the three samples examined in this study at $n_{2DEG} = 2.9\times10^{11}~\text{cm}^{-2}$. \textit{Z} is the distance from top barrier-dielectric interface. (c) Strain energy density for different quantum well thicknesses as a function of the underlying GBL lattice parameter. The dashed vertical lines mark previously reported GBL terminal layer lattice parameters, assuming total relaxation~\cite{Shabani2014APL,Hatke2017APL,Dempsey2025PRM}. The $\text{In}_{0.875}\text{Al}_{0.125}\text{As}$ GBL used in this study is indicated by the dotted vertical line.
    }
    \label{S-P_solver}
  \end{minipage}
\end{figure*}

\section{Molecular Beam Epitaxy Growth}

Three wafers were grown in a VEECO GEN 930 molecular beam epitaxy (MBE) system with quantum well thicknesses of 10, 12, and 14 ~nm. A schematic of the full heterostructure design is presented in Fig.~1(a). All samples were grown on (001) oriented Fe-doped semi-insulating InP substrates. To preserve the integrity of the MBE vacuum, the substrates were prepared in an ancillary ultra-high vacuum (UHV) chamber by heating to \qty{300}{\degreeCelsius}, outgassing residual water and volatile organic compounds. 

In the MBE chamber sample surface reconstructions were monitored in real-time using reflection high-energy electron diffraction (RHEED), while temperature was monitored via optical pyrometry and band-edge thermometry. Upon loading into the MBE reactor, substrates were heated under an As$_4$ overpressure until reaching ~10 degrees below the thermal oxide desorption temperature of InP. The As$_4$ overpressure was maintained at a constant beam equivalent pressure of $3\times10^{-5}$ Torr. While monitoring RHEED, the As$_4$ overpressure was interrupted by a shutter in front of the source for several seconds, during which the surface reconstruction was observed to switch from a 2$\times$3 pattern to a metal-rich 4$\times$2 pattern. After observing this transition, the As$_4$ overpressure was reapplied, yielding a 2$\times$4 reconstruction. This oxide removal technique minimizes the time spent under the metal-rich phase, thus reducing degradation of the substrate surface prior to growth. After removal of the oxide, the samples were cooled to $470^\circ\text{C}$ before initiating  growth of a lattice-matched 100~nm $\text{In}_{0.52}\text{Al}_{0.48}\text{As}$ layer and the $\text{In}_{0.58}\text{Ga}_{0.42}\text{As}$/$\text{In}_{0.47}\text{Al}_{0.53}\text{As}$ strained-layer superlattice (SLSL).

After completion of the SLSL, the substrate temperature was further reduced to \qty{250}{\degreeCelsius} prior to growth of the 1250~nm InAlAs step-graded buffer. Instead of utilizing a linear grading profile as described extensively in the literature~\cite{Shabani2014APL,Hatke2017APL,Dempsey2025PRM}, the Al mole fraction was decreased from 48\% to 12.5\% with a decaying exponential profile, resulting in very large changes in Al-concentration and lattice parameter during the initial layers of the GBL, and smaller changes towards the final layers, thereby minimizing the formation of misfit dislocations and associated threading dislocations near the active region of the heterostructure. Substrate temperature was precisely controlled during GBL growth by automated feedback on the band-edge thermometry, minimizing sample-to-sample temperature variations. The substrate temperature was held constant at \qty{250}{\degreeCelsius} for the initial 550~nm of the GBL, after which it was linearly ramped from \qty{250}{\degreeCelsius} to \qty{350}{\degreeCelsius} over the remaining 700~nm. Higher growth temperatures increase dislocation glide velocity, and therefore reduce threading dislocation density (TDD) in the active region by promoting dislocation annihilation in the GBL~\cite{Kim2022NRL,George1987}. However, higher growth temperatures also amplify the crosshatch surface morphology typical of metamorphic graded buffers by increasing mobility of the group III adatoms~\cite{Rovaris2019}. The choice of temperature profile during growth of the GBL can therefore be tailored to the required specifications of a given heterostructure, and provides a potential future avenue for optimization. 

Upon completion of the GBL, the substrate temperature was increased to \qty{450}{\degreeCelsius} for growth of the active region, which consists of a 25~nm $\text{In}_{0.875}\text{Al}_{0.125}\text{As}$ lower barrier, an InAs quantum well, and a 120~nm $\text{In}_{0.875}\text{Al}_{0.125}\text{As}$ top barrier. Three different quantum well thicknesses were grown, as detailed in Table~\ref{Table_1}.

\section{Results}

\subsection{Schr\"{o}dinger-Poisson and Strain Calculations}
To better understand the effect of widening the quantum well in these structures, we modeled Samples A through C using the NextNano$^{3}$\cite{Nextnano} self-consistent Schr\"{o}dinger-Poisson solver. The conduction band edge and electron density distribution is plotted for the three samples in Fig.~\ref{S-P_solver}(b). The degree of confinement of the electrons in the quantum well was also calculated at a density of $n_{2DEG} = 2.9\times10^{11}~\text{cm}^{-2}$, which is close to the density at which maximum mobility was observed for the three samples. As expected, confinement in the quantum well increases with increasing well thickness. Additionally, the areal strain energy density was calculated as a function of the GBL lattice parameter for the three quantum well widths examined in this work, as shown in Fig.~\ref{S-P_solver}(c). 

\subsection{Structural Characterization}

\begin{figure*}[!t]
  \centering
  \begin{minipage}{\textwidth}
    \centering
    \includegraphics[width=\textwidth]{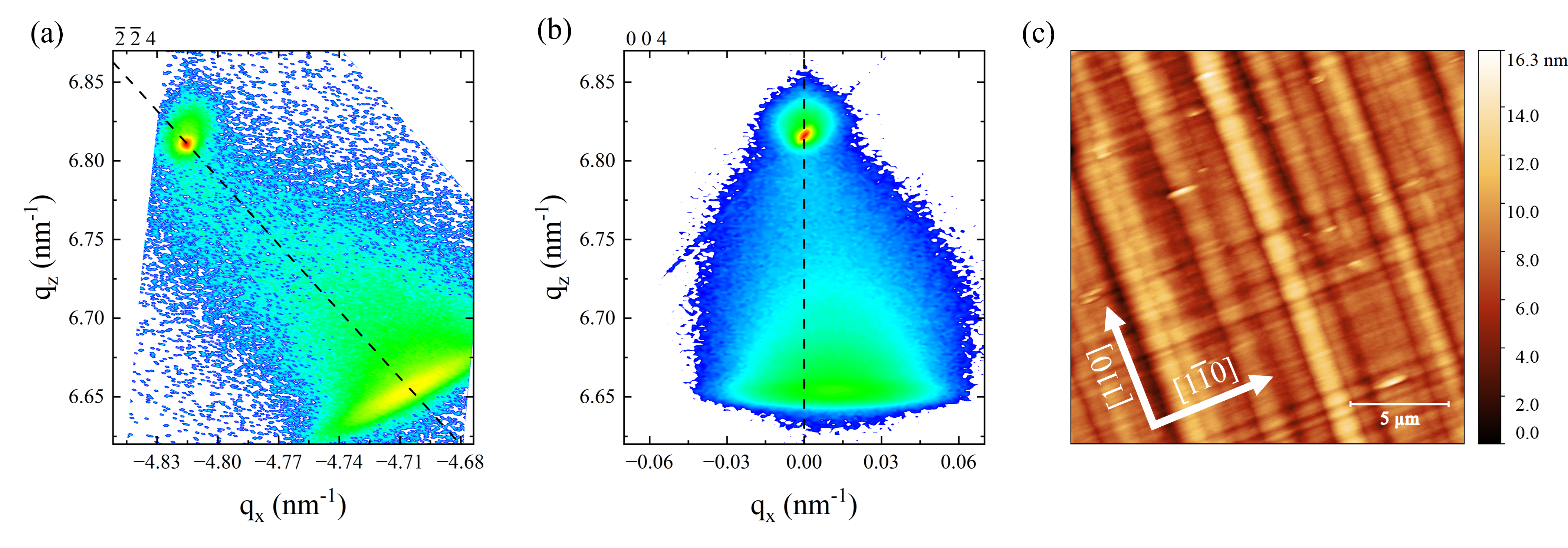}
    \caption{Reciprocal space map of Sample A around the asymmetric $(\bar{2} \bar{2}4)$ reflection in (a) and the symmetric $(004)$ reflection in (b). The diagonal dashed line in (a) is the line of total relaxation, while the vertical dashed line in (b) marks $q_x = 0$. While the intensity maximum appear to track along the relaxation line in the $(\bar{2} \bar{2}4)$ reflection, one must correct for the macroscopic tilt shown by the deviation from $q_x = 0$ in the $(004)$ reflection. With this correction, a small amount of residual tensile strain is observed for the samples in this study. (c) A $20\times20~\upmu\text{m} ^2$ AFM scan of Sample A, exhibiting the hallmark cross-hatch morphology of compositionally graded buffers.
    }
    \label{RSM and AFM}
  \end{minipage}
\end{figure*}

To characterize the strain state and alloy composition of the GBL, a Malvern Panalytical Empyrean X-ray diffractometer was used to collect high-resolution reciprocal space maps (RSM) around the symmetric $(004)$ and asymmetric $(\bar{2} \bar{2}4)$ reflections, as shown in Fig.~\ref{RSM and AFM}(a) and (b), respectively. The symmetric $(004)$ peak can be used to extract the out-of-plane lattice parameter, as well as the macroscopic tilt of the epitaxial layers. The asymmetric $(\bar{2} \bar{2}4)$ peak can be used to extract the in-plane lattice parameter in addition to the out-of-plane lattice parameter, but this information is convoluted with macroscopic layer tilt. It is possible to deconvolve the peak shifts caused by macroscopic tilt from those caused by strain using basic geometric corrections, as demonstrated by Chauveau \textit{et al}.~\cite{Chauveau2003JAP}. After applying these corrections, the GBL was found to have a residual tensile strain of $\epsilon_{\sslash} = 0.16 \%$. While it is often desirable to produce a GBL with no residual strain, the use of strain-compensating layers in III-V systems to extend the effective critical thickness of the quantum well has been demonstrated numerous times~\cite{Tansu2001,Choi1999,Dempsey2025PRM}. Our data suggests that the residual strain in the GBL is advantageous to achieving thicker coherently strained quantum wells. 

In addition to X-ray measurements, we collected atomic force microscopy (AFM) scans of all three samples to characterize surface morphology. AFM of Sample A is shown in Fig.~\ref{RSM and AFM}(c). While the samples exhibit the typical crosshatch morphology of compositionally graded buffers, we did not see any obvious signs of plastic relaxation in the active region, as was demonstrated in the work of Lei {\it et al.} where they intentionally push the InAs quantum well past its critical thickness limit~\cite{Lei2026}. The two-dimensional root-mean-square (RMS) roughness, $R_{q}^{2D}$, is 1.97~nm in Sample A, and a large anisotropy of the crosshatch was observed between the two major crystallographic directions, $[110]$ and $[1\bar{1}0]$. This anisotropy was quantified by taking the linear RMS roughness along these directions. For sample A, $R_q=1.9~\text{nm}$ along $[1\bar{1}0]$, while $R_q=0.6~\text{nm}$ along $[110]$. We further characterized the anisotropy by comparing the crosshatch wavelength, $\lambda$, by taking a Fast Fourier Transform (FFT) along the two directions. Again in Sample A, $\lambda=3.3~\upmu\text{m}$ along $[1\bar{1}0]$, while $\lambda=2.2~\upmu\text{m}$ along $[110]$.

\begin{figure*}[!t]
    \centering
    \includegraphics[width=1.15\textwidth]{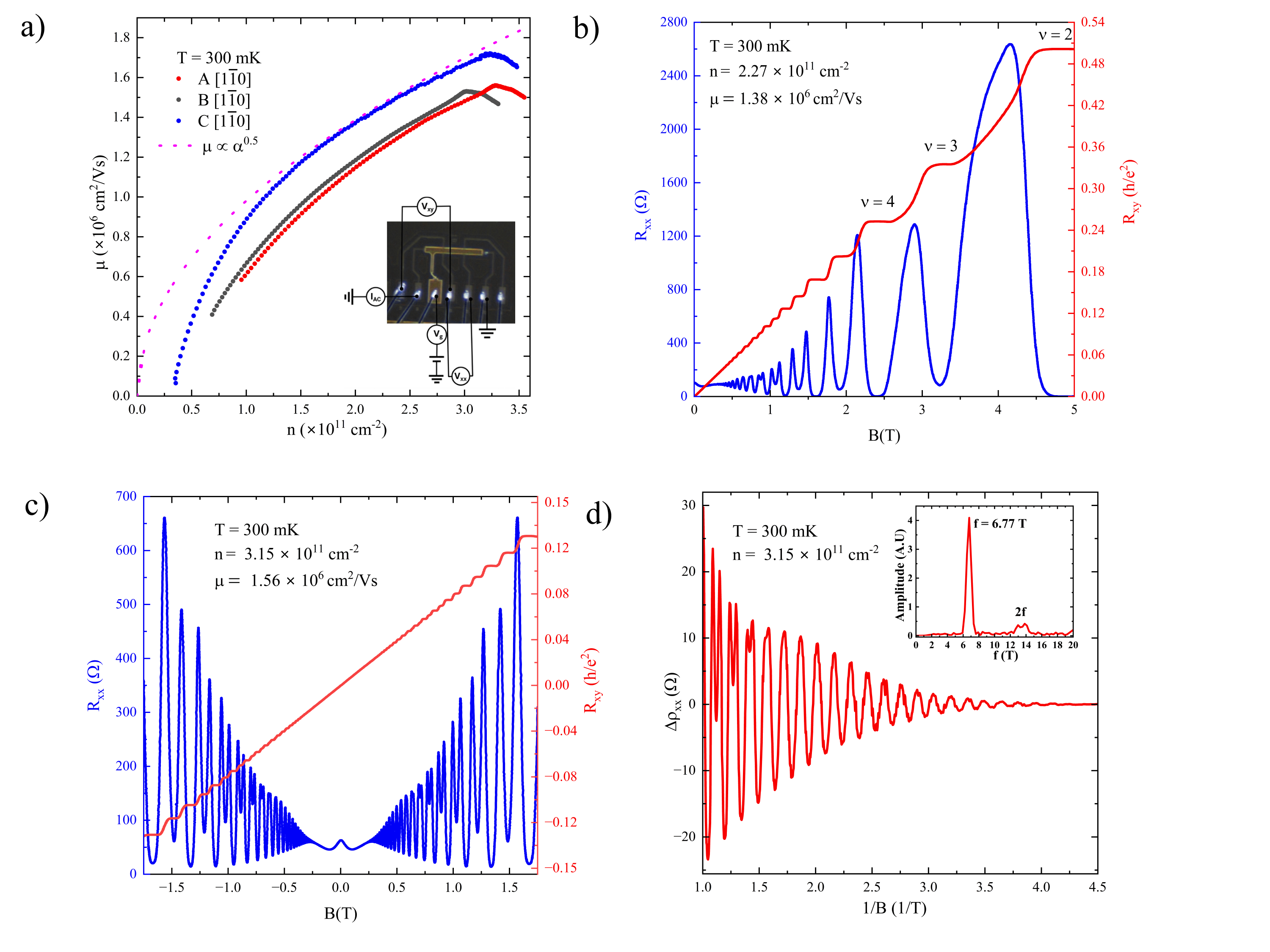}
    \caption{(a) Mobility vs carrier density for Samples A, B and C measured along  [1$\bar{1}$0] crystallographic direction. The dotted line represents the power law $\mu \propto n^{\alpha}$, with ${\alpha}$ = 0.5. The peak mobility of $1.72\times10^6~\mathrm{cm^2\,V^{-1}\,s^{-1}}$ is measured in Sample C (14~nm quantum well) at  $n=3.24\times10^{11}\,\text{cm}^{-2}$. Inset: An optical microscope image of one of the fabricated Hall bars, annotated with a schematic of the measurement circuit.  (b) Magnetotransport for the 14~nm wide quantum well at  $n=2.27\times10^{11}\,\text{cm}^{-2}$ measured at $T =300~\text{mK}$ with the integer filling factors labeled. c) Low-field R$_{xx}$ and R$_{xy}$ for 14~nm quantum well at $n=3.15\times10^{11}\,\text{cm}^{-2}$ measured at $T =300~\text{mK}$ showing the onset of Shubnikov–de Haas oscillations at around $\pm$ 0.2~T. d) Oscillatory part of the longitudinal resistivity for $n=3.15 \times10^{11}$cm$^{-2}$ at low field after subtracting a smooth polynomial background as a function of $\frac{1}{B}$. Inset: FFT of the longitudinal resistivity. The dominant peak ($f$) indicates the density of the single conducting channel; the second harmonic at 
    $2f$ is associated with the onset of spin splitting.}
    \label{magnetotransport}
\end{figure*}

\subsection{Device Fabrication and Measurement}

To facilitate electronic transport measurements at different 2DEG densities, top-gated Hall bars with nominal widths of $120~\upmu\text{m}$ were fabricated on all samples. Hall bars were oriented along both the [1$1$0] and [1$\bar{1}$0] crystallographic directions. Mesas were defined for the Hall bars by a standard wet etch process. Following the etch, Ar ion milling was used to remove the native oxide followed by deposition of Ti/Al Ohmic contacts by electron-beam evaporation. The top gate dielectric is 11~nm of halfnia grown via a standard thermal ALD process. Finally, a Ti/Au gate is electron-beam evaporated and patterned via a lift-off process to cover the Hall bar.

Transport measurements are performed at $T =300~\text{mK}$, with a magnetic field up to 5~T, using low frequency AC excitation of 100 nA to measure the longitudinal voltage ($V_{xx}$) and transverse voltage ($V_{xy}$) across the sample with the magnetic field in perpendicular direction as shown in inset of  Fig.~\ref{magnetotransport}(a).

\subsection{Mobility versus well width and 2DEG density}
In order to understand the 2DEG density dependence on gate voltage, $R_{xy}$ was first extracted at $B = 0.1~\text{T}$ as the function of gate voltage for each Hall bar. This data facilitates study of the dependence of mobility on 2DEG density and comparison of transport in the different quantum wells at comparable 2DEG densities. 

We then examine the quantum well width dependence of the peak mobility. Here we focus on transport along the $[1\bar{1}0]$ direction, as peak mobility is slightly higher in this direction due to anisotropy associated with the crosshatch morphology. For Sample C with a well width of 14~nm, the peak mobility was $\mu=1.72 \times10^6~\mathrm{cm^2\,V^{-1}\,s^{-1}}$ at $n=3.24\times10^{11}~\text{cm}^{-2}$. This is the highest mobility observed in our study. As depicted in Table \ref{Table_1}, the mobility of all the samples is $\geq 1.5 \times 10^{6} \mathrm{cm^2}/\mathrm{Vs}$ with a weak diminution in peak mobility for quantum wells of thickness 12~nm and 10~nm. Evidently the difference in interface roughness scattering between 10~nm and 12~nm wide quantum wells is insufficient to produce a significant change in mobility. More dramatic reduction is expected for well widths below 10~nm.

To understand the scattering mechanisms that limit mobility in our current generation of structures, we plot $\mu$ versus $n_{2DEG}$ for all three quantum well widths in Fig.~\ref{magnetotransport}(a). The mobility rises with increasing 2DEG density over the range of $0.5\times10^{11}$cm$^{-2}$ to approximately $3.0\times10^{11}$cm$^{-2}$. At approximately $n_{2DEG}=3.0\times10^{11}$cm$^{-2}$, the mobility peaks and then reduces for further increase in density. This behavior is associated with the onset of population of the second electric subband in the quantum well. Note that the 10~nm and 12~nm quantum wells behave in a similar manner to the 14~nm quantum albeit with reduced peak mobility. 

In Fig.~\ref{magnetotransport}(a) we also plot the functional form $\mu \propto \alpha^{0.5}$ for comparison with experimental data. The exponent $\alpha \approx $ 0.5 indicates that the dominant scattering mechanism is uniformly distributed background charged impurity scattering~\cite{DasSarma2013PRB}. We see that at 2DEG density lower than $1.0\times10^{11}$cm$^{-2}$, the mobility falls more rapidly than $\mu \propto \alpha^{0.5}$, indicating a transition to a more strongly localized transport regime. A similar value of $\alpha$ was extracted in Hatke \textit{et al.} for InAs quantum wells on InP substrates where the cladding barriers were In$_{0.75}$Ga$_{0.25}$As~\cite{Hatke2017APL} and Dempsey \textit{et al.}~\cite{Dempsey2025PRM} where the In concentration in cladding barriers were reduced to 0.72. The improvement in peak mobility observed in this generation of heterostructures may be attributed to reduction of the background charged impurity density and reduction of the impact of interface roughness scattering achieved in the wider InAs quantum wells. It is interesting to compare previous results to those presented here. To make this comparison, the proper figure of merit is $\mu / n_{2DEG}^{0.5}$. For the best sample of Hatke {\it et al.}~\cite{Hatke2017APL}, this ratio is 0.44 while for Sample C of the current study this ratio is 0.96. Our current data indicates that we have reduced scattering by a factor of $\geq2$ over the previous state of the art.

\subsection{Magnetotransport Measurements}
 Magnetotransport for Sample C at carrier density $2.27\times10^{11}~\text{cm}^{-2}$ at $T =300~\text{mK}$ and in magnetic field to $B = 5~\text{T}$ is shown in Fig.~\ref{magnetotransport}(b). The filling factor $\nu$ of several integer quantum Hall states is indicated. Figure \ref{magnetotransport}(c) focuses on the low B-field region at $n_{2DEG}=3.15 \times10^{11}~\text{cm}^{-2}$.  We observe the onset of Shubnikov–de Haas oscillation (SdH) at $B \approx0.2$~T. At very small magnetic field in the weak-localization regime, when spin-orbit coupling is negligible, time-reversal symmetry allows pairs of closed electron trajectories traversed in opposite directions to interfere constructively. This constructive interference enhances the probability of coherent backscattering, leading to an increase in the longitudinal resistance, $R_{xx}$, at zero magnetic field. When a perpendicular magnetic field is applied, an additional magnetic phase is accumulated between the time-reversed paths. This phase difference suppresses the constructive interference and reduces the backscattering probability \cite{farzaneh2024observing}. As a result, $R_{xx}$ decreases with increasing magnetic field, producing a negative magnetoresistance characteristic of weak localization \cite{golub2005weak}, as evident in Fig.~\ref{magnetotransport}(c). Figure~\ref{magnetotransport}(d) presents the oscillatory part of the longitudinal resistivity for Sample C after subtracting the slowly varying background as function of $\frac{1}{B}$. The inset shows the  FFT of the longitudinal resistivity. The prominent fundamental peak at 6.77 T indicates there is a single conducting band while the smaller signal at $2f$ marks the onset of spin splitting. Using the Onsager relation, $n = \frac{|e| f}{h}$ \cite{onsager1952interpretation}, the carrier density is determined to be 3.22 $\times$ 10$^{11}$ cm$^{-2}$ in agreement with the carrier concentration extracted from the slope of the Hall resistance 3.15 $\times$ 10$^{11}$ cm$^{-2}$, confirming the absence of parallel conduction.

\section{Conclusion}
Through optimization of the GBL, we have demonstrated InAs quantum wells with thicknesses up to 14~nm integrated onto InP semi-insulating substrates. The use of a convex grading profile allowed the GBL to reach a larger lattice parameter and lower Al alloy composition than would be possible with a linear GBL and compositional step-back, without introducing a parallel conduction channel. Further, we found through X-ray analysis that we do have a small amount of tensile residual strain in these GBLs, which allowed for thick QWs through a strain-compensation effect without introducing additional alloy scattering by use of higher Ga-content cladding barriers. Finally, low-temperature electronic transport in these samples showed substantial improvement over previously reported InAs QW on InP, with the 14 nm sample demonstrating a peak mobility $\geq1.7\times10^6~\mathrm{cm^2\,V^{-1}\,s^{-1}}$ at a 2DEG density \mbox{$\leq3.3\times10^{11}~\text{cm}^{-2}$.}

\begin{acknowledgements}
This work was supported by Microsoft Quantum.
\end{acknowledgements}

\section*{AUTHOR DECLARATIONS} 

\subsection*{Conflict of Interest} The authors have no conflicts to disclose. 

\subsection*{Author Contributions} Tyler Lindemann: Conceptualization; Investigation; Methodology; Formal analysis; Visualization; Writing -- original draft; Writing -- review \& editing. Rojila Ghimire: Investigation; Formal analysis; Visualization; Writing -- original draft; Writing -- review \& editing. Alejandro Alcaraz Ramirez: Investigation; Formal analysis; Writing -- review \& editing.  Ahmad Azizimanesh: Investigation; Writing -- review \& editing. Sergei Gronin: Methodology; Writing -- review \& editing. Ray Kallaher: Investigation; Writing -- review \& editing. Michael J. Manfra: Conceptualization; Supervision; Project administration; Funding acquisition;  Writing -- original draft; Writing -- review \& editing. 

\section*{DATA AVAILABILITY} The data that support the findings of this study are available from the corresponding author upon reasonable request.

\bibliographystyle{aipnum4-2}
\bibliography{citedpaper}

@article{Chauveau2003JAP,
  author    = {Chauveau, J.-M. and Androussi, Y. and Lefebvre, A. and Di Persio, J. and Cordier, Y.},
  title     = {Indium content measurements in metamorphic high electron mobility transistor structures by combination of {X}-ray reciprocal space mapping and transmission electron microscopy},
  journal   = {Journal of Applied Physics},
  volume    = {93},
  number    = {7},
  pages     = {4219--4225},
  year      = {2003},
  doi       = {10.1063/1.1544074},
  publisher = {American Institute of Physics}
}

@article{Hatke2017APL,
  author  = {Hatke, A. T. and Wang, T. and Thomas, C. and Gardner, G. C. and Manfra, M. J.},
  title   = {Mobility in excess of $10^{6}~\mathrm{cm}^{2}/\mathrm{V}\,\mathrm{s}$ in {InAs} quantum wells grown on lattice mismatched {InP} substrates},
  journal = {Applied Physics Letters},
  volume  = {111},
  pages   = {142106},
  year    = {2017},
  doi     = {10.1063/1.4993784}
}

@article{Dempsey2025PRM,
  title     = {Effects of strain compensation on electron mobilities in {InAs} quantum wells grown on {InP}(001)},
  author    = {Dempsey, C. P. and Dong, J. T. and Villar Rodriguez, I. and Gul, Y. and Chatterjee, S. and Pendharkar, M. and Holmes, S. N. and Pepper, M. and Palmstr{\o}m, C. J.},
  journal   = {Physical Review Materials},
  volume    = {9},
  number    = {5},
  pages     = {054607},
  year      = {2025},
  month     = may,
  publisher = {American Physical Society},
  doi       = {10.1103/PhysRevMaterials.9.054607},
  url       = {https://doi.org/10.1103/PhysRevMaterials.9.054607}
}

@incollection{Ayers2015,
  author    = {Ayers, John E.},
  title     = {Low-temperature and metamorphic buffer layers},
  booktitle = {Handbook of Crystal Growth: Thin Films and Epitaxy},
  editor    = {Kuech, Thomas F.},
  edition   = {2nd},
  volume    = {III},
  chapter   = {25},
  pages     = {1007--1056},
  publisher = {Elsevier},
  year      = {2015},
  isbn      = {978-0-444-63304-0},
  doi       = {10.1016/B978-0-444-63304-0.00025-1},
  url       = {https://doi.org/10.1016/B978-0-444-63304-0.00025-1}
}

@article{Shabani2014APL,
  author  = {Shabani, J. and McFadden, A. P. and Shojaei, B. and Palmstrom, C. J.},
  title   = {Gating of high-mobility {InAs} metamorphic heterostructures},
  journal = {Applied Physics Letters},
  volume  = {105},
  pages   = {262105},
  year    = {2014},
  doi     = {10.1063/1.4905370}
}

@misc{Nextnano,
  author = {{nextnano GmbH}},
  note   = {{nextnano software}, https://www.nextnano.com}
}

@article{Tansu2001,
  author    = {Tansu, Nelson and Mawst, Luke J.},
  title     = {High-performance strain-compensated {InGaAs}-{GaAsP}-{GaAs} ($\lambda = 1.17~\mu\mathrm{m}$) quantum well diode lasers},
  journal   = {IEEE Photonics Technology Letters},
  volume    = {13},
  number    = {3},
  pages     = {179--181},
  year      = {2001},
  doi       = {10.1109/68.914313},
}

@article{Choi1999,
  author    = {Choi, Won-Jin and Dapkus, P. Daniel and Jewell, Jack L.},
  title     = {$1.2~\mu\mathrm{m}$ {GaAsP}/{InGaAs} strain-compensated single-quantum-well diode laser on {GaAs} using metal-organic chemical vapor deposition},
  journal   = {IEEE Photonics Technology Letters},
  volume    = {11},
  number    = {12},
  pages     = {1572--1574},
  year      = {1999},
  doi       = {10.1109/68.806850},
}

@article{DasSarma2013PRB,
  author  = {Das Sarma, S. and Hwang, E. H.},
  title   = {Universal density scaling of disorder-limited low-temperature conductivity in high-mobility two-dimensional systems},
  journal = {Physical Review B},
  volume  = {88},
  pages   = {035439},
  year    = {2013},
  doi     = {10.1103/PhysRevB.88.035439}
}

@article{Kim2022NRL,
  author    = {Kim, HoSung and Geum, Dae-Myeong and Ko, Young-Ho and Han, Won-Seok},
  title     = {Effects of high-temperature growth of dislocation filter layers in {GaAs}-on-{Si}},
  journal   = {Nanoscale Research Letters},
  volume    = {17},
  pages     = {126},
  year      = {2022},
  doi       = {10.1186/s11671-022-03762-9},
  url       = {https://www.ncbi.nlm.nih.gov/pmc/articles/PMC9763523/}
}

@article{George1987,
  author    = {George, A. and Rabier, J.},
  title     = {Dislocations and plasticity in semiconductors. {I}---Dislocation structures and dynamics},
  journal   = {Revue de Physique Appliqu\'{e}e},
  volume    = {22},
  number    = {9},
  pages     = {941--966},
  year      = {1987},
  doi       = {10.1051/rphysap:01987002209094100}
}

@article{Rovaris2019,
  author    = {Rovaris, Fabrizio and Zoellner, Marvin H. and Zaumseil, Peter and Marzegalli, Anna and Di Gaspare, Luciana and De Seta, Monica and Schroeder, Thomas and Storck, Peter and Schwalb, Georg and Capellini, Giovanni and Montalenti, Francesco},
  title     = {Dynamics of crosshatch patterns in heteroepitaxy},
  journal   = {Physical Review B},
  volume    = {100},
  number    = {8},
  pages     = {085307},
  year      = {2019},
  doi       = {10.1103/PhysRevB.100.085307}
}

@article{Lei2026,
  title   = {Impact of layer structure and strain on morphology and electronic properties of {InAs} quantum wells on {InP} (001)},
  author  = {Lei, Zijin and Wu, Yuze and Reichl, Christian and Fält, Stefan and Wegscheider, Werner},
  journal = {arXiv preprint arXiv:2603.07303},
  year    = {2026},
  eprint  = {2603.07303},
  archivePrefix = {arXiv},
  primaryClass = {cond-mat.mtrl-sci},
  url     = {https://arxiv.org/abs/2603.07303}
}

@article{Kroemer2004,
  author    = {H. Kroemer},
  title     = {The 6.1~{\AA} family ({InAs}, {GaSb}, {AlSb}) and its heterostructures: A selective review},
  journal   = {Physica E: Low-dimensional Systems and Nanostructures},
  volume    = {20},
  number    = {3--4},
  pages     = {196--203},
  year      = {2004},
  issn      = {1386-9477}
}

@article{Thomas2018,
  author  = {C. Thomas and A. T. Hatke and A. Tuaz and R. Kallaher and T. Wu and T. Wang and R. E. Diaz and G. C. Gardner and M. A. Capano and M. J. Manfra},
  title   = {High-mobility {InAs} {2DEGs} on {GaSb} substrates: A platform for mesoscopic quantum transport},
  journal = {Physical Review Materials},
  volume  = {2},
  pages   = {104602},
  year    = {2018},
  doi     = {10.1103/PhysRevMaterials.2.104602}
}

@article{Tuttle1989,
  author  = {Gary Tuttle and Herbert Kroemer and John H. English},
  title   = {Electron concentrations and mobilities in {AlSb/InAs/AlSb} quantum wells},
  journal = {Journal of Applied Physics},
  volume  = {65},
  number  = {12},
  pages   = {5239--5242},
  year    = {1989},
  doi     = {10.1063/1.343167}
}

@article{farzaneh2024observing,
  title={Observing magnetoanisotropic weak antilocalization in near-surface quantum wells},
  author={Farzaneh, SM and Hatefipour, Mehdi and Schiela, William F and Lotfizadeh, Neda and Yu, Peng and Elfeky, Bassel Heiba and Strickland, William M and Matos-Abiague, Alex and Shabani, Javad},
  journal={Physical Review Research},
  volume={6},
  number={1},
  pages={013039},
  year={2024},
  publisher={APS}
}

@article{golub2005weak,
  title={Weak antilocalization in high-mobility two-dimensional systems},
  author={Golub, LE},
  journal={Physical Review B},
  volume={71},
  number={23},
  pages={235310},
  year={2005},
  publisher={APS}
}

@article{benali2022metamorphic,
title = {Metamorphic {InAs/InGaAs} {QWs} with electron mobilities exceeding $7 \times 10^{5}~\mathrm{cm}^{2}/\mathrm{V}\,\mathrm{s}$},
  author={Benali, A and Rajak, P and Ciancio, R and Plaisier, JR and Heun, S and Biasiol, G},
  journal={Journal of Crystal Growth},
  volume={593},
  pages={126768},
  year={2022},
  publisher={Elsevier}
}

@article{onsager1952interpretation,
  title={Interpretation of the de {Haas}-van {Alphen} effect},
  author={Onsager, Lars},
  journal={The London, Edinburgh, and Dublin Philosophical Magazine and Journal of Science},
  volume={43},
  number={344},
  pages={1006--1008},
  year={1952},
  publisher={Taylor \& Francis}
}

\end{document}